\documentclass[fleqn,10pt]{wlscirep}
\usepackage{graphicx}
\title{Navigation by anomalous random walks on complex networks}

\author[1,*]{Tongfeng Weng}
\author[2]{Jie Zhang}
\author[1]{Moein Khajehnejad}
\author[3,4]{Michael Small}
\author[1]{Rui Zheng}
\author[1,+]{Pan Hui}
\affil[1]{HKUST-DT System and Media Laboratory, Hong Kong University of Science and Technology, HongKong}
\affil[2]{Centre for Computational Systems Biology, Fudan University, China}
\affil[3]{The University of Western Australia, Crawley, WA 6009, Australia}
\affil[4]{Mineral Resources Flagship, CSIRO, Kensington, WA, Australia}

\affil[*]{wtongfeng2006@163.com}
\affil[+]{panhui@cse.ust.hk}

\begin{abstract}
Anomalous random walks having long-range jumps are a critical branch of dynamical processes on networks, which can model a number of search and transport processes. However, traditional measurements based on mean first passage time are not useful as they fail to characterize the cost associated with each jump. Here we introduce a new concept of mean first traverse distance (MFTD) to characterize anomalous random walks that represents the expected traverse distance taken by walkers searching from source node to target node, and we provide a procedure for calculating the MFTD between two nodes. We use L\'{e}vy walks on networks as an example, and demonstrate that the proposed approach can unravel the interplay between diffusion dynamics of L\'{e}vy walks and the underlying network structure. Interestingly, applying our framework to the famous PageRank search, we can explain why its damping factor empirically chosen to be around 0.85. The framework for analyzing anomalous random walks on complex networks offers a new useful paradigm to understand the dynamics of anomalous diffusion processes, and provides a unified scheme to characterize search and transport processes on networks. 
\end{abstract}
\begin{document}

\flushbottom
\maketitle
\thispagestyle{empty}

\section*{Introduction}

Complex networks are ubiquitous in the real world ranging from sociology to biology and technology \cite{Barabasi2012}. Going beyond the interesting topological properties, quantifying the impact of structural organization of networks on transport processes has become one of the most important topics. As a paradigmatic transport process, random walks on complex networks have been intensively studied \cite{JDNoh2004,SCondamin2007,SHwang2012,Perra2012,YLin2014}. A variety of measurements including mean first passage time (MFPT) \cite{JDNoh2004}, first passage time \cite{SHwang2012}, and average trapping time \cite{YLin2014} have been proposed, providing a comprehensive characterization of random walks on networks. Moreover, these studies also facilitate our understanding of diverse dynamical processes on networks including epidemic spreading \cite{ZMYang2012}, synchronization \cite{PSSkardal2014}, and transportation \cite{GLi2010}.

However, for random walks, the walker is confined only to the neighbourhood of a node in each jump, which cannot model some real situations \cite{FDPatti2015}, and also impedes search and transport efficiency on networks \cite{SHwang2012}. This limitation is circumvented by the model of L\'{e}vy walks in natural condition \cite{GMViswanathan2010,DRaichlen2014}. Recently, intensive attention has been devoted to anomalous random walks on networks, such as L\'{e}vy walks \cite{APRiascos2012,APRiascos2014}, traditional web surfing \cite{ANLangville2006}, and even electric signals transmitted in brain networks \cite{FDPatti2015}. One striking feature of anomalous random walks is having the long-range hopping (i.e., the walker can hop to far away nodes not directly connected to its current position). In fact, the occurrence of long-range hopping is frequently encountered in our life. For example, we usually communicate with people socially close to us, but also occasionally with those that are unconnected \cite{APRiascos2014}. Analogously, when doing web surfing, one usually proceeds by following the hyperlinks but casually may open a new tab to look for the related topic \cite{FDPatti2015}. Although it is widely agreed that anomalous random walks represent an important branch of search and transport processes on networks, how to characterize anomalous random walks and specifically, to uncover the interplay between their dynamics and the underlying network structure has not been addressed. Traditional measurements like the mean first passage time neglect the difference between the cost associated with the nearest-neighbor jump and the long-range hopping, therefore cannot properly characterize anomalous random walks on networks. 

In this paper, we propose the mean first traverse distance that represents the expected traverse distance required by a walker moving from a source node to a target node. Importantly, this allows the cost associated with the hopping to be taken into account in the characterization of  
anomalous random walks; this therefore overcomes the problems of traditional measurements adopted in general random walks. We obtain analytically the MFTD and the global MFTD on arbitrary networks. Results on L\'{e}vy walks demonstrate that these measurements can effectively characterize the relationship between network structure and anomalous random walks. Interestingly, when applied to the PageRank search, we demonstrate that the optimal damping factor occurs at around 0.85 in real web networks which is consistent with our empirical finding. The new metric enables effective characterization of dynamics of anomalous random walks on networks, which promises more efficient search and transport processes on networks.

\section*{Results}

\textbf{The MFTD of anomalous random walks} We start from an undirected network consisting of $N$ nodes. The connectivity of nodes is fully described by a symmetric adjacency matrix $A$, whose entry $a_{ij}=1$ (0) if nodes $i$ and $j$ are (not) connected. For anomalous random walks, at each time step, the walker jumps from current node $i$ to node $j$ with a nonzero transition probability $p_{ij}$ regardless of the connection profile of  node $i$. Take L\'{e}vy walks on networks for example, the transition probability is defined as $p_{ij}={d_{ij}^{-\alpha}}/{\sum_{k}d_{ik}^{-\alpha}}$, where $\alpha$ is the tuning exponent lying in the interval $0\leq{\alpha}<\infty$ and $d_{ij}$ is the shortest path length between nodes $i$ and $j$ \cite{APRiascos2012}. To characterize anomalous random walks, we propose a concept of the MFTD $l_{ij}$, which is the expected distance taken by a walker to first reach node $j$ starting from node $i$. Intuitively, the traverse distance in one-step jump is shorter for a walker when nodes are directly connected, while this distance tends to be larger for indirectly linked nodes. Inspired by the empirical findings that the lengths of links usually obey a power law distribution \cite{Li2011}, we adopt the power function $c_{ij}=d_{ij}^{\beta}$ to describe the effective distance of one-step jump, where $\beta$ named the cost exponent is a nonnegative value. In this situation, if the first step of the walk is to node $j$, the expected traverse distance required is $d_{ij}^{\beta}$; if it is to some other node $k$, the expected traverse distance becomes $l_{kj}$ plus $d_{ik}^{\beta}$ for the previous step already taken. Thus, we have
\begin{equation}
\ l_{ij}=p_{ij}d_{ij}^{\beta}+\sum_{k\neq{j}}p_{ik}(l_{kj}+d_{ik}^{\beta}).
\label{101}
\end{equation}
Using the Markov chains theory \cite{Grinstead2006,Kemeny1960}, the MFTD $l_{ij}$ of a anomalous random walk (see appendices) becomes
\begin{equation}
\ l_{ij}=T_{ij}\sum_{k}\left(\sum_{m}p_{km}d_{km}^{\beta}\right)w_{k}+\sum_{k}(z_{ik}-z_{jk})\left(\sum_{m}p_{km}d_{km}^{\beta}\right),
\label{102}
\end{equation}
where $w_{k}$ is the $k$th component of the stationary distribution of the anomalous random walk, $T_{ij}$ is the MFPT from node $i$ to node $j$, and $z_{ij}$ is an element of the fundamental matrix $Z=(I-P+W)^{-1}$. Specifically, when $\beta=0$, the effective distances of one-step jump are same (i.e., $c_{ij}=1$). In this situation, it is easy to verify that the MFTD $l_{ij}$ reduces to the MFPT $T_{ij}$, which means that our paradigm can incorporate the commonly used MFPT as a special case. To further evaluate the search efficiency of an anomalous walker, we calculate the global MFTD $\langle{L}\rangle$ by averaging Eq.~(\ref{102}) over all pairs of source and target nodes, that is,
\begin{equation}
\ \langle{L}\rangle=\frac{1}{N(N-1)}\sum_{i}^{N}\sum_{j}^{N}l_{ij}.
\label{32}
\end{equation}
Plugging Eq.~(\ref{102}) into Eq.~(\ref{32}), we have
\begin{equation}
\ \langle{L}\rangle=\langle{T}\rangle{\sum_{i}\left(\sum_{j}p_{ij}d_{ij}^{\beta}\right)w_{i}},
\label{33}
\end{equation}
where $\langle{T}\rangle$ is the average of MFPTs over all pairs of nodes in the networks (see appendices). Here, $\langle{L}\rangle$ quantifies the ability of the anomalous walker to search and transport at the global scale on the network. In this context, smaller $\langle{L}\rangle$ represents a more effective way of achieving mobility. In the following we will demonstrate how these measurements can effectively characterize diverse anomalous random walks on networks.

\begin{figure}[!htb]
\centering
\includegraphics[width=1\textwidth,height=0.32\textheight]{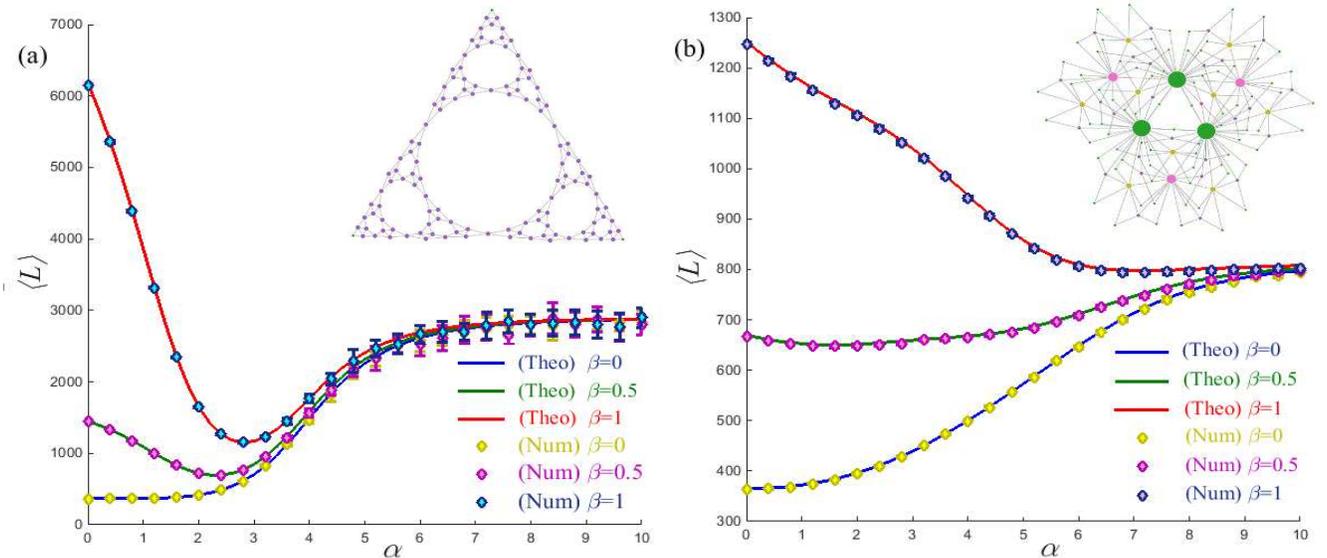}
\caption {The global MFTD $\langle{L}\rangle$ as a function of $\alpha$, for L\'{e}vy walks on (a) the planar Sierpi\'{n}ski gasket and (b) the (1,2) flower model with the same size $N=366$ nodes and $\beta=0, 0.5, 1$, respectively. Symbols represent the values of $\langle{L}\rangle$ found numerically, while solid lines correspond to the theoretical prediction of Eq.~(\ref{7}). Error bars represent the mean first traverse distance $\langle{L}\rangle$ over 20 tests and each test is averaged over the ensemble of 50,000 independent runs.}\label{f1}
\end{figure}

\textbf{The MFTD scheme for characterizing L\'{e}vy walks} We first address a specific anomalous random walk --- L\'{e}vy walks on networks. A L\'{e}vy walk exerts a power-law transition probability with the distance given by $p_{ij}={d_{ij}^{-\alpha}}/{\sum_{k}d_{ik}^{-\alpha}}$. Clearly, the tuning exponent $\alpha$ plays an important role in controlling the trade off between short-range and long-range jumping in one step, which in turn fully determines the behaviors of the L\'{e}vy walk. Specially, when $\alpha$ is very small, the walker visits all nodes with approximately equivalent probability. In contrast, the walker possibly only hop to the nearest neighbors at an extremely large $\alpha$. In this context, the L\'{e}vy walk degenerates to the generic random walk \cite{JDNoh2004}. Using the balance condition, the stationary distribution of the L\'{e}vy walk can be expressed as
\begin{equation}
\ w_{i}=\frac{\sum_{k}d_{ik}^{-\alpha}}{\sum_{i}\sum_{j}d_{ij}^{-\alpha}}.
\label{5}
 \end{equation}
Inserting the above equation and the transition probability into Eq.~(\ref{102}) yields
\begin{equation}
\ l_{ij}=\frac{z_{jj}-z_{ij}}{\sum_{m}d_{jm}^{-\alpha}}\sum_{i}\sum_{j}d_{ij}^{\beta-\alpha}+\sum_{k}(z_{ik}-z_{jk})\frac{\sum_{m}d_{km}^{\beta-\alpha}}{\sum_{m}d_{km}^{-\alpha}}.
\label{6}
\end{equation}
Similar calculation applied to Eq.~(\ref{33}), the global MFTD $\langle{L}\rangle$ of L\'{e}vy walks reads
\begin{equation}
\ \langle{L}\rangle=\langle{T}\rangle{\frac{\sum_{i}\sum_{j}d_{ij}^{\beta-\alpha}}{\sum_{i}\sum_{j}d_{ij}^{-\alpha}}}.
\label{7}
\end{equation}
To test the validity of Eq.~(\ref{7}), we report both the numerical and theoretical results of the global MFTD for L\'{e}vy walks taking place in planar Sierpi\'{n}ski gasket \cite{JJKozak2002} and the (1,2)-flower model \cite{HDRozenfeld2007}. These two networks are typical hierarchical nets having the same number of nodes and edges but exhibiting apparently distinct structure organizations, which can favor us to explore how the network structure influences the behavior of a L\'{e}vy walk directly. To achieve the numerical results, we compute the traverse distance required for a walker to travel from a source node to a target node chosen randomly and average over the ensemble of 50,000 independent runs for each test. Fig.~\ref{f1} shows an excellent agreement between numerics and Eq.~(\ref{7}) for the different cost exponents $\beta$. In particular, when $\beta=0$, the minimum of $\langle{L}\rangle$ occurs at $\alpha=0$ regardless of the network structures, which reproduces the previous results based on the MFPT \cite{YLin2013}. However, this result is unreasonable in practice without considering the distinct costs induced by the nearest-neighborhood jumps and the long-range hops. In contrast, we find that when $\beta>0$, the profiles of different network organizations show clearly distinct behaviors. Specially, the profiles of the planar Sierpi\'{n}ski gasket display a clear minimum in the medium range $\alpha$, which minimizes the search distance, (i.e., the global mean first traverse distance). However, such behavior is absent for the (1,2)-flower model for $\beta>0$, where they present a clearly monotonous tendency, see Fig.~\ref{f1} (b). Such difference can be intuitively explained when referring to their topological properties. Specially, the Sierpi\'{n}ski gasket is a fractal network without the ``small-world'' property \cite{JJKozak2002}, see its topological structure in Fig.~\ref{f1} (a). In contrast, the (1,2)-flower network has the ``small-world'' feature and the ``scale-free'' characteristics \cite{HDRozenfeld2007}, as shown in Fig.~\ref{f1} (b). Meanwhile, we also notice that, when $\alpha$ is large, the transition probability of the L\'{e}vy walk degenerates to a generic random walk. Thereby, all curves approach a fixed value for $\alpha>9$, see in Fig.~\ref{f1} (a) and (b), as expected.

\begin{figure}[!htb]
\centering
\includegraphics[width=1\textwidth,height=0.65\textheight]{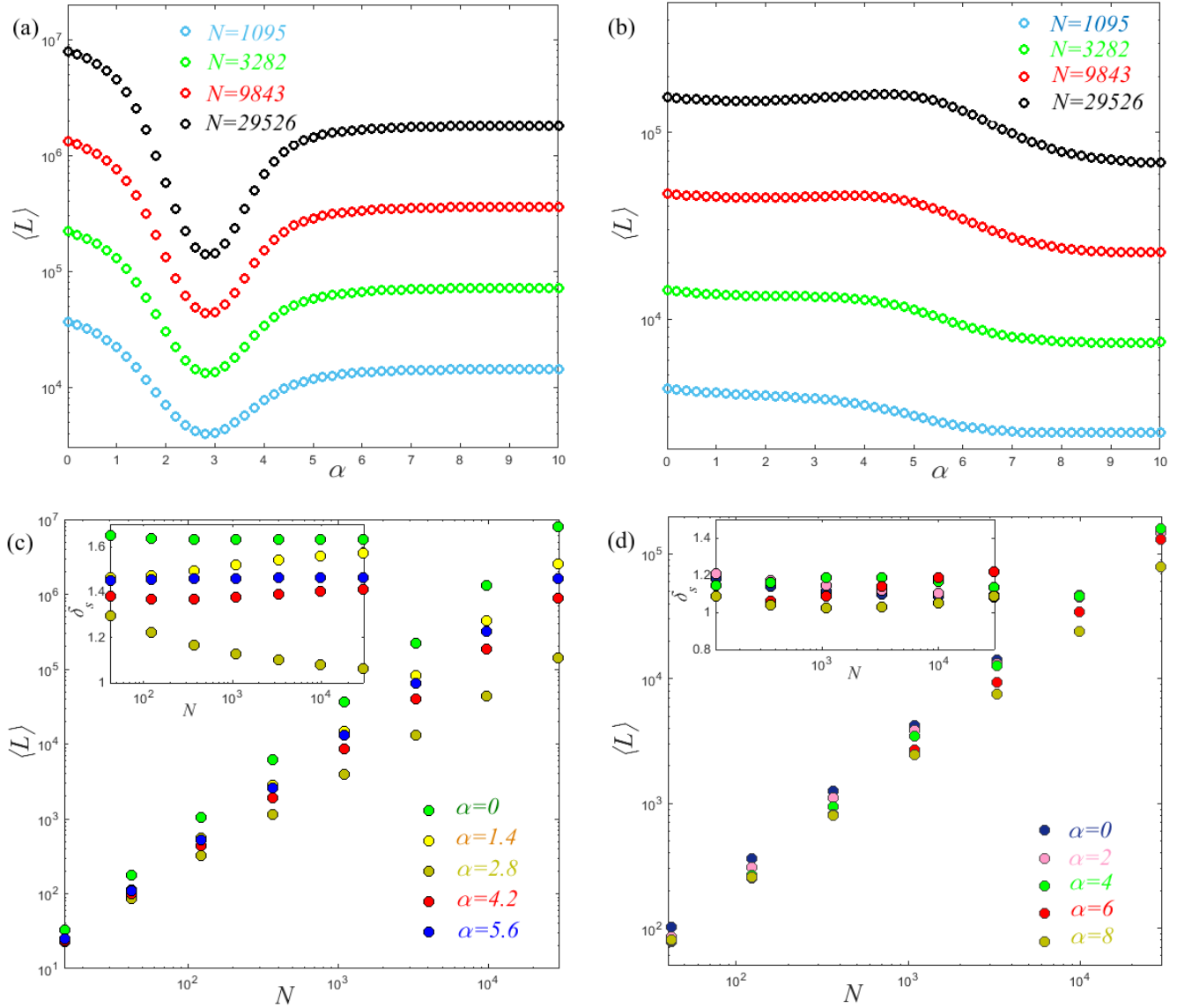}
\caption {The global MFTD $\langle{L}\rangle$ as a function of $\alpha$ for L\'{e}vy walks on (a) the planar Sierpi\'{n}ski gasket and (b) the (1,2) flower model over different network sizes $N$. The behaviors of $\langle{L}\rangle$ versus $N$ for different exponents $\alpha$ on the planar Sierpi\'{n}ski gasket (c) and the (1,2) flower model (d). In the insets, we show the plots of the successive slopes $\delta_{s}$ obtained from $ln\langle{L}\rangle$ versus $lnN$. Note that here we set the cost exponent $\beta=1$.}\label{f5}
\end{figure}

To further demonstrate the difference induced by network structure, we observe the size effect on the global MFTD $\langle{L}\rangle$ of the planar Sierpi\'{n}ski gasket and the (1,2)-flower model. We find that the profiles of each network present the same tendency for different network sizes $N$, see Fig.~\ref{f5} (a) and (b). Interestingly, the result presented in Fig.~\ref{f5} (a) clearly shows the presence of a minimum $\langle{L}\rangle$ for different network sizes at the same exponent $\alpha=2.8$. The way in which $\langle{L}\rangle$ scales with network size $N$ on the planar Sierpi\'{n}ski gasket seems to follow rather different behaviors depending on the tuning exponent $\alpha$. Specially, when $\alpha{\neq}2.8$, the global MFTD $\langle{L}\rangle$ follows a power law with network size $N$, see in Fig.~\ref{f5} (c). It is supported by observing the almost invariant values of the successive slopes $\delta_{s}$ obtained from $ln\langle{L}\rangle$ versus $lnN$, as shown in the inset of Fig.~\ref{f5} (c). Conversely, for $\alpha=2.8$, the successive slopes $\delta_{s}$ present a clearly decreasing tendency. However, for the (1,2)-flower model, the $\langle{L}\rangle$ follows approximately a power law with network size $N$, see in Fig.~\ref{f5} (d). Note that here we choose the cost exponent $\beta=1$ for convenience. However, such behavior of $\langle{L}\rangle$ versus $N$ is general for an arbitrary cost exponent $\beta$. 

\begin{figure}[!htb]
\centering
\includegraphics[width=0.95\textwidth,height=0.8\textheight]{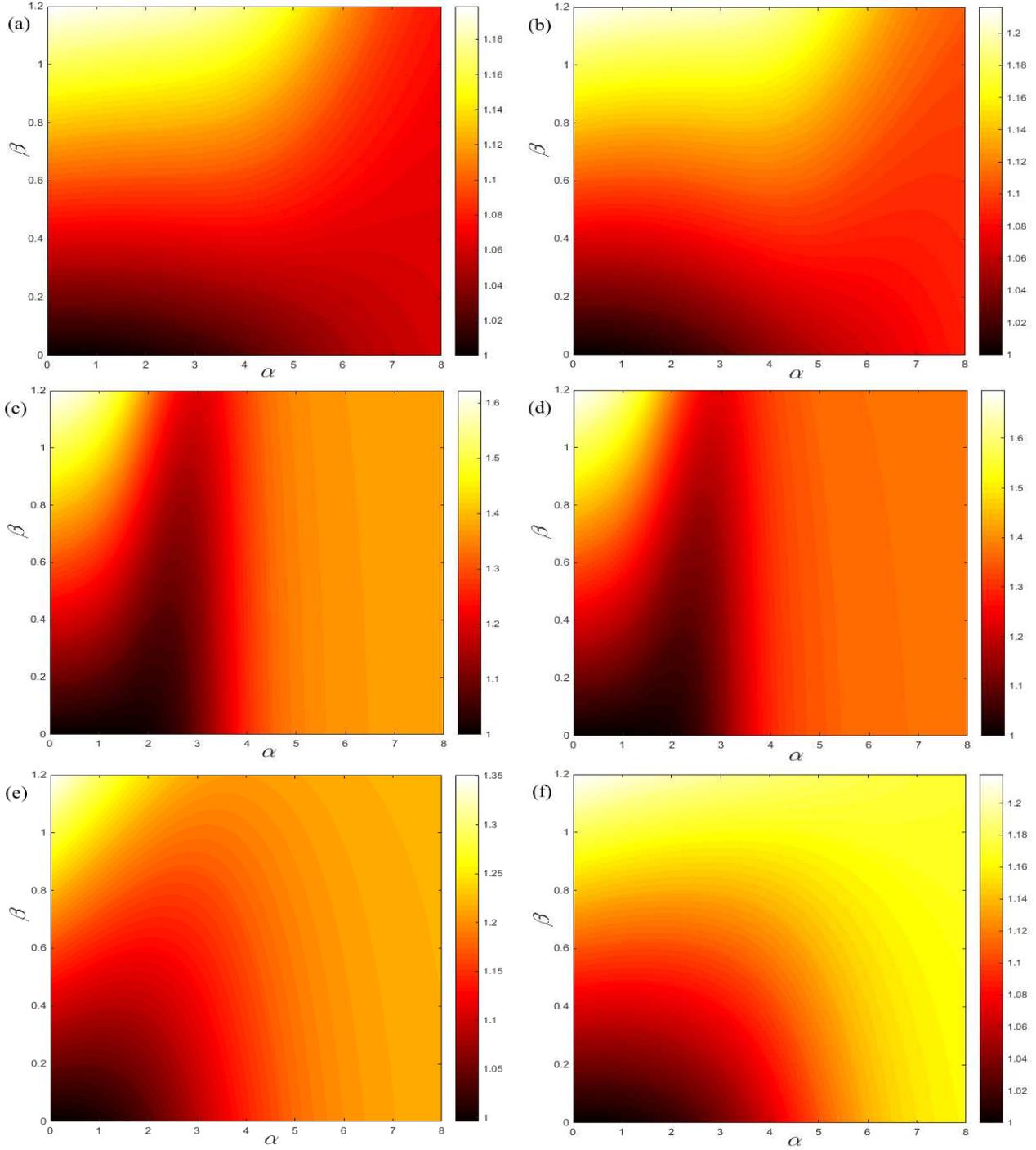}
\caption {The measurement $log_{N}\langle{L}\rangle$ in the $(\alpha,\beta)$ parameter plane of (a) the BA model, (b) the (1,2)-flower model, (c) planar Sierpi\'{n}ski gasket, (d) the (4,5)-flower model, (e) the ``Dolphin'' network \cite{DLusseau2003}, and (f) the e-mail network \cite{RGuimera2003}.}\label{f2}
\end{figure}

Clearly, from Eq.~(\ref{7}), the cost exponent $\beta$ plays an important role in controlling the search efficiency for L\'{e}vy walks. In order to explore how the optimal search efficiency of a L\'{e}vy walk changes with respect to the cost exponent $\beta$, we investigate the interplay between $\beta$ and $\alpha$ for various networks including three synthetic models (the Barab\'{a}si-Albert (BA) model \cite{ABarabasi1999}, the planar Sierpi\'{n}ski gasket \cite{JJKozak2002}, and the (u,v)-flower model \cite{HDRozenfeld2007}) and two real networks (the ``Dolphin'' network \cite{DLusseau2003} and an e-mail network \cite{RGuimera2003}). Here, for a fair comparison, we calculate the measurement $log_{N}\langle{L}\rangle$ in the $(\alpha,\beta)$ plane for eliminating the size effect of networks. Generally, regions with smaller $log_{N}\langle{L}\rangle$ indicate an efficient way of search and transport based on L\'{e}vy walks. Fig.~\ref{f2} shows contour maps of $log_{N}\langle{L}\rangle$ in the $(\alpha,\beta)$ plane computed for these selected networks. Interestingly, we find that distinct network structures lead to different patterns in the corresponding $(\alpha,\beta)$ plane. Specifically,  the $(\alpha,\beta)$ planes generated from networks having the ``small-world'' characteristics, such as the BA model and the (1,2)-flower model, demonstrate an ``estuary'' pattern, implying that L\'{e}vy walks are not the optimal way to search when $\beta>0.4$. In contrast, typical fractal networks without the ``small-world'' property, for example, the planar Sierpi\'{n}ski gasket and the (4,5)-flower model, result in a striking ``flame''  in the $(\alpha,\beta)$ planes, suggesting that there exists an optimal tuning exponent $\alpha$, which minimizes the traverse distance for a broad range of cost exponents $\beta$. However, none of these patterns match the ones found in the Dolphin network and the e-mail network, whose $(\alpha,\beta)$ planes show ``rippled'' features, meaning that the optimal exponent $\alpha$ gradually increases with the cost exponent $\beta$. The $(\alpha,\beta)$ plane uncovers the relationship between network structure and the behavior of L\'{e}vy walks, which provides information to help designing more effective search strategies and transport mechanisms in different environments. 

\begin{figure}[!htb]
\centering
\includegraphics[width=1\textwidth,height=0.65\textheight]{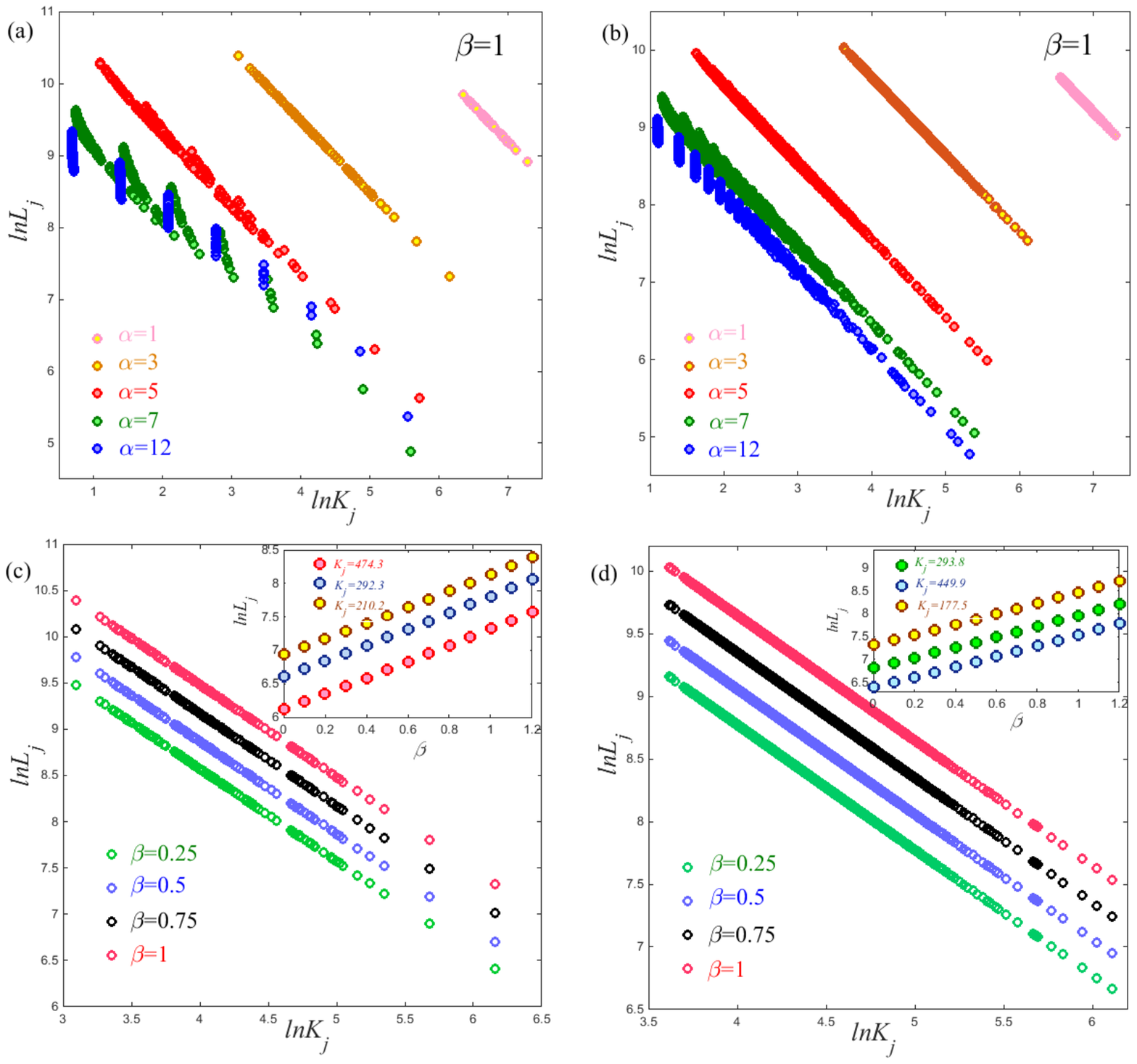}
\caption {Plots of $lnL_{j}$ versus $lnK_{j}$ are presented for (a) the (1,2)-flower model and (b) the BA model with network size $N=3282$. The same plots with respect to different cost exponents $\beta$ for (c) the (1,2)-flower mode and (d) the BA model under the cost exponent $\alpha=3$. In the inset, we show $lnL_{j}$ versus $\beta$ for different trapping nodes $j$. Note that here the values of $lnL_{j}$ are calculated based on the Eq.~(\ref{27}).}\label{f3}
\end{figure}

Furthermore, we follow the spirit of the MFPT, and extract more statistics from the MFTD. Here, we introduce the average trapping distance (ATD) defined as follows:
 \begin{equation}
\  L_{j}=\frac{1}{1-w_{j}}\sum_{m=1}^{N}w_{m}l_{mj}.
 \label{27}
 \end{equation}
The ATD $L_{j}$ quantifies the mean of MFTD $l_{mj}$ to the trap node $j$, taken over all starting points with the stationary distribution. Submitting the results of Eqs.~(\ref{5}) and (\ref{6}) into Eq.~(\ref{27}) yields (see appendices)
 \begin{equation}
\  L_{j}\approx{\frac{z_{jj}}{K_{j}}\sum_{i}\sum_{j}d_{ij}^{\beta-\alpha}},
 \label{28}
 \end{equation}
where $K_{j}=\sum_{m}d_{jm}^{-\alpha}$ named the long-range degree of node $j$ \cite{APRiascos2012}. Specifically, when $\alpha$ is small, the diagonal values of $Z$ are almost same. In this context, a clear scaling behavior emerges such that $L_{j}\sim{K_{j}^{-1}}$ regardless of the underlying network structure. This is supported by observing the plots of $lnL_{j}$ vs $lnK_{j}$ shown in Fig.~\ref{f3} (a) and (b). With an increase of $\alpha$, the slope of $lnL_{j}$ versus $lnK_{j}$ gradually decreases and finally asymptotically approaches to that of random walks as described in Ref. \citen{SHwang2012}. Results demonstrate the important role of $\alpha$ in shaping the ATD. Meanwhile, from Eq.~(\ref{28}), it is easy to verify that the relationship between $lnL_{j}$ and $ln K_{j}$ does not depend on the cost exponent $\beta$. So, the profiles present a similar tendency for different cost exponents $\beta$ as illustrated in Fig.~\ref{f3} (c) and (d). We further find a linear relationship between $lnL_{j}$ and $\beta$, when fixing the tapping position $j$ and the tuning exponent $\alpha$, see the insets in Fig.~\ref{f3} (c) and (d). The results are consistent with our theoretical prediction of the relationship $lnL_{j}\sim{C\beta}$, where $C$ is a constant value related to the fractal dimension of a given network (see appendices). 

\begin{figure}[!htb]
\centering
\includegraphics[width=1\textwidth,height=0.65\textheight]{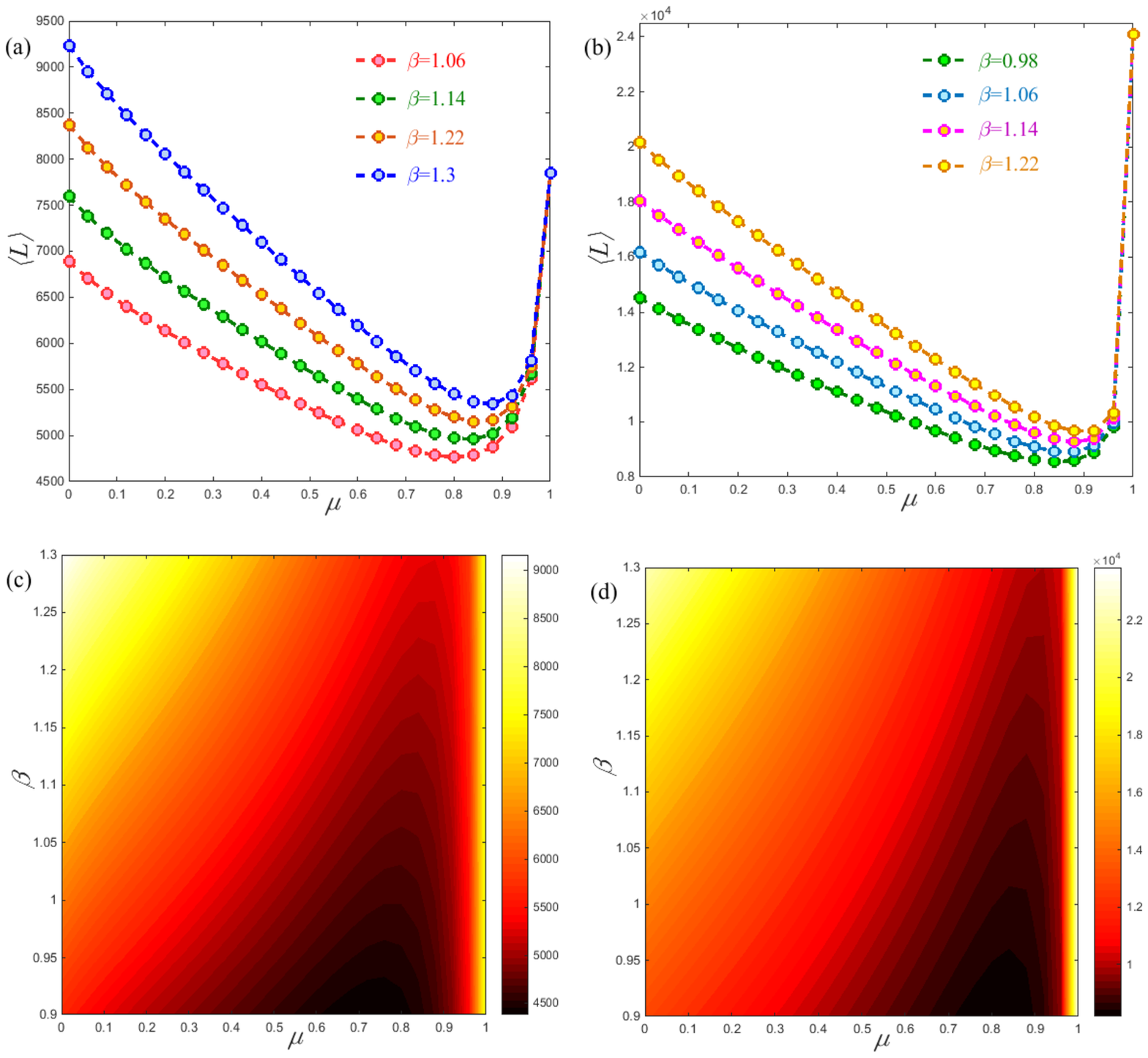}
\caption {The global MFTD $\langle{L}\rangle$ as a function of the damping factor $\mu$, for the PageRank search on (a) web-Stanford \cite{JLeskovec2009} and (b) Ego-Facebook \cite{JLeskovec2012}. Symbols correspond to the theoretical prediction of Eq.~(\ref{33}). The global MFTD $\langle{L}\rangle$ in the $(\beta,\mu)$ parameter plane of (c) web-Stanford and (d) Ego-Facebook. Note that the web-Stanford network used here is a subgraph extracted from the original one for computation convenience with $N=2004$.}\label{f4}
\end{figure}

\textbf{The optimal condition of the PageRank search based on the MFTD theory} We finally apply the MFTD theory to characterize the famous PageRank search \cite{ANLangville2006}. The PageRank search is widely used to compute the relevance of web pages. The transition probability $p_{ij}$ of the PageRank search is
 \begin{equation}
\ p_{ij}={\mu}\frac{a_{ij}}{k_{i}}+(1-\mu){\frac{1}{N}}
\label{31}
\end{equation}
where $k_{i}=\sum_{l}a_{il}$ is the degree of node $i$ and $\mu$ is the damping factor lying in the range $[0,1]$. We investigate the global MFTD  $\langle{L}\rangle$ for the PageRank search on two real networks (web-Stanford \cite{JLeskovec2009} and Ego-Facebook \cite{JLeskovec2012}). The results presented in Fig.~{\ref{f4}} (a) and (b) indicate the existence of a minimum $\langle{L}\rangle$ for different cost exponents $\beta$ at the same value of the damping factor $\mu{\approx}0.85$, where optimal search is achieved. This is further supported by observing the contour maps of the $(\mu,\beta)$ plane, where for $\mu{\approx}0.85$, the global MFTD $\langle{L}\rangle$ is near its minimum value for a very broad range of $\beta$, see in Fig.~{\ref{f4}} (c) and (d). This can explain why the ad hoc damping factor of the PageRank search is suggested to be set around 0.85. Moreover, we notice that the minimum $\langle{L}\rangle$ of the PageRank search is much smaller than that of generic random walks (i.e., $\mu=1$), which in some extent demonstrates the advantage of taking the PageRank search instead of generic random walks. 

\section*{Discussion}

In summary, we have introduced the concept of the MFTD, a measure that takes into account of the cost of jumps in anomalous random walks, therefore is particularly suited to capture the interplay between the diffusion dynamics of anomalous random walks and underlying network structures. We obtain an exact expression for the MFTD and the global MFTD of anomalous random walks on complex networks. We show that our paradigm provides a unified scheme to characterize diffusion processes on networks, which incorporates the commonly used MFPT as a special case. 

We then demonstrate the effectiveness of these measures by applying them to L\'{e}vy walks. We find that distinct network structures result in different patterns in the $(\alpha,\beta)$ planes, which explores the effect of the cost exponent $\beta$ on behaviors of L\'{e}vy walks with respect to network structure. Moreover, when addressing the trapping problem of L\'{e}vy walks, we find that its behavior only depends on the tuning exponent $\alpha$ irrespective of the cost exponent $\beta$. In particularly, when $\alpha$ is smaller, it presents a uniformly scaling feature regardless of network structure. These findings enrich our understanding of interplay between dynamics of  L\'{e}vy walks and network structure. To implement L\'{e}vy walks, we need to compute all shortest paths of a network which involves high computational costs for large networks. In practice, one can use several excellent algorithms such as the preprocessing algorithm \cite{CSommer2014}, which is one of possible solution for this problem. Nonetheless, the results show that for a broad range of the cost exponent $\beta$, the global MFTD $\langle{L}\rangle$ of L\'{e}vy walks is much smaller than that of generic random walks, which demonstrates the efficient for search and transport based on L\'{e}vy walks.

Finally, application to the famous PageRank search shows that the empirical damping factor is optimal at 0.85 for the cost exponent lying in the range $[0.9,1.3]$, at which individuals can optimize search in traverse distance. It is suggested that the time required for opening a new tab is approximately equivalent to that of following the hyperlinks for several turns for a related topic search. Thereby, in practice, the damping factor of the PageRank search is chosen around 0.85. Overall, our findings offer a new framework to understand the diffusion dynamics of anomalous random walks on complex networks. 

\section*{Appendices}

\textbf{The analytic expression of mean first traverse distance} We follow the derivation of MFPT in Ref. \citen{Grinstead2006} to calculate the MFTD on networks. We consider an arbitrary finite network consisting of $N$ nodes. The connectivity is represented by the adjacency matrix $A$, whose entries $a_{ij}=1$ (or 0) if there is (not) a link from nodes $i$ to $j$. Let $D$ denote the distance matrix with elements $d_{ij}$ representing the shortest path length from node $i$ to node $j$. In the process of anomalous random walks, at each step, the walker starting from node $i$ arrives to node $j$ with a non-zero transition probability $p_{ij}$ regardless of the connectivity between nodes $i$ and $j$. If the first step of the walk is to node $j$, the expected traverse distance required is $d_{ij}^{\beta}$; if it is to some other node $k$, the expected traverse distance becomes $l_{kj}$ plus $d_{ik}^{\beta}$ for the previous step already taken. Thus, we obtain
\begin{equation}
\ l_{ij}=p_{ij}d_{ij}^{\beta}+\sum_{k\neq{j}}p_{ik}(l_{kj}+d_{ik}^{\beta}),
\label{51}
\end{equation}
where $l_{ij}$ is the mean first traverse distance from node $i$ to node $j$. Since $l_{jj}=0$, Eq.~(\ref{51}) can be rewritten as
\begin{equation}
\ l_{ij}=\sum_{m}{p_{im}d_{im}^{\beta}}+\sum_{k}p_{ik}l_{kj}.
\label{52}
\end{equation}
Let $r_{i}$ denote the mean first return distance to node $i$ starting from node $i$. In the same manner, $r_{i}$ can be represented as
 \begin{equation}
\ r_{i}=\sum_{k}p_{ik}(l_{ki}+d_{ik}^{\beta}).
\label{53}
\end{equation}
Combining Eq.~(\ref{52}) and Eq.~(\ref{53}) together, we obtain the relation
\begin{equation}
\ (I-P)L=C-R,
\label{54}
\end{equation}
where $I$ denotes the identity matrix, and
\begin{equation}
L=\left(\begin{array}{cccc}
 l_{11}&l_{12}& \cdots &l_{1n}\\
 l_{21}&l_{22}&\cdots&l_{2n}\\
 \vdots&\vdots&\vdots&\vdots\\
 l_{n1}&l_{2n}&\cdots&l_{nn}
 \end{array} \right),
 \end{equation}
 
 \begin{equation}
C=\left(\begin{array}{cccc}
 \sum_{k}p_{1k}d_{1k}^{\beta}&\sum_{k}p_{1k}d_{1k}^{\beta}
 & \cdots&\sum_{k}p_{1k}d_{1k}^{\beta}\\
 \sum_{k}p_{2k}d_{2k}^{\beta}& \sum_{k}p_{2k}d_{2k}^{\beta}& \cdots&\sum_{k}p_{2k}d_{2k}^{\beta}\\
 \vdots&\vdots&\vdots&\vdots\\
 \sum_{k}p_{Nk}d_{Nk}^{\beta}& \sum_{k}p_{Nk}d_{Nk}^{\beta}&\cdots& \sum_{k}p_{Nk}d_{Nk}^{\beta}
 \end{array} \right),
 \end{equation}

\begin{equation}
R=\left(\begin{array}{cccc}
 r_{1}&0& \cdots &0\\
 0&r_{2}&\cdots&0\\
 \vdots&\vdots&\vdots&\vdots\\
 0&0&\cdots&r_{n}
 \end{array} \right).
 \end{equation}
 Multiplying both sides of Eq.~(\ref{54}) by the matrix $W=\left(\begin{array}{cccc}
 w_{1}&w_{2}&\cdots&w_{N}\\
w_{1}&w_{2}&\cdots&w_{N}\\
 \vdots&\vdots&\vdots&\vdots\\
 w_{1}&w_{2}&\cdots&w_{N}
 \end{array} \right)$ with the element $w_{i}$ being the $i$th component of the stationary distribution, and using the fact that
  \begin{equation}
\ W(I-P)=0
\label{55}
\end{equation}
gives
 \begin{equation}
\ WC-WR=0.
\label{56}
\end{equation}
From Eq.~(\ref{56}), the mean first return distance $r_{i}$ reads
 \begin{equation}
\ r_{i}=\frac{\sum_{k}\left(\sum_{m}p_{km}d_{km}^{\beta}\right)w_{k}}{w_{i}}.
\label{57}
\end{equation}
Since the matrix $(I-P+W)$ has an inverse \cite{Grinstead2006}, we denote $Z=(I-P+W)^{-1}$. Multiplying 
both sides of Equation (\ref{54}) by $Z$ and using the fact that
  \begin{equation}
\ I-W=Z(I-P)
\label{58}
\end{equation}
gives
 \begin{equation}
\ L=ZC-ZR+WL.
\label{59}
\end{equation}
From the above equation, $l_{ij}$ and $l_{jj}$ can be expressed as
\begin{equation}
\ l_{ij}= \sum_{k}z_{ik}\left(\sum_{m}p_{km}d_{km}^{\beta}\right)-z_{ij}r_{j}+(wL)_{j}
\label{10}
\end{equation}
and
\begin{equation}
\ l_{jj}= \sum_{k}z_{jk}\left(\sum_{m}p_{km}d_{km}^{\beta}\right)-z_{jj}r_{j}+(wL)_{j}.
\label{11}
\end{equation}
Since $l_{jj}=0$ and using Eq.~(\ref{57}), one has
\begin{equation}
\ l_{ij}=T_{ij}\sum_{k}\left(\sum_{m}p_{km}d_{km}^{\beta}\right)w_{k}+\sum_{k}(z_{ik}-z_{jk})\left(\sum_{m}p_{km}d_{km}^{\beta}\right),
\label{12}
\end{equation}
where $T_{ij}=\frac{z_{jj}-z_{ij}}{w_{j}}$ is the mean first passage time.

 \textbf{The analytic expression of global mean first traverse distance} 
 To further evaluate the search efficiency based on anomalous random walks, we introduce the global mean first traverse distance defined as
\begin{equation}
\ \langle{L}\rangle=\frac{1}{N(N-1)}\sum_{i}\sum_{j}{l_{ij}}.
\label{13}
\end{equation}
Plugging Eq.~(\ref{12}) into Eq.~(\ref{13}), we obtain
\begin{equation}
\ \langle{L}\rangle=\langle{T}\rangle{\sum_{k}\left(\sum_{m}p_{km}d_{km}^{\beta}\right)w_{k}}+\frac{1}{N(N-1)}\sum_{i}\sum_{j}\sum_{k}(z_{ik}-z_{jk})\left(\sum_{m}p_{km}d_{km}^{\beta}\right),
\label{14}
\end{equation}
where $\langle{T}\rangle=\frac{1}{N(N-1)}\sum_{i}\sum_{j}{T_{ij}}$ is the global mean first passage time. Since column vectors of the matrix $C$ are the same, the column vectors of the matrix $ZC$ is also the same. Then, the last term of Eq.~(\ref{14}) will vanish due to that
 \begin{equation}
\ \sum\sum{(ZC-(ZC)^{T}})=0,
\label{15}
\end{equation}
 where the matrix $(ZC)^{T}$ represents the transpose of matrix $ZC$. So, the expression for $\langle{L}\rangle$ is reduced to 
\begin{equation}
\ \langle{L}\rangle=\langle{T}\rangle{\sum_{k}\left(\sum_{m}p_{km}d_{km}^{\beta}\right)w_{k}}.
\label{16}
\end{equation}

\textbf{The analytic expression of average trapping distance for L\'{e}vy walks} We now study the trapping problem for L\'{e}vy walks at an arbitrarily given node. Let $L_{j}$ be the average trapping distance, which is the mean of MFTD $L_{ij}$ to the trap node $j$, taken over the stationary distribution defined as follows:
 \begin{equation}
\  L_{j}=\frac{1}{1-w_{j}}\sum_{i=1}^{N}w_{i}l_{ij}.
 \label{17}
 \end{equation}
 Substituting the expression of $l_{ij}$ in Eq.~(\ref{6}) and $w_{j}$ in Eq.~(\ref{5}) into Eq.~(\ref{17}) gives
  \begin{equation}
\  L_{j}=\frac{1}{1-w_{j}}\sum_{i=1}^{N}w_{i}\left(\frac{z_{jj}-z_{ij}}{w_{j}}\frac{\sum_{i}\sum_{j}d_{ij}^{\beta-\alpha}}{\sum_{i}\sum_{j}d_{ij}^{-\alpha}}\right)+\frac{1}{1-w_{j}}\sum_{i=1}^{N}w_{i}\left(\sum_{k}(z_{ik}-z_{jk})\left(\frac{\sum_{m}d_{km}^{\beta-\alpha}}{\sum_{m}d_{km}^{-\alpha}}\right)\right).
 \label{18}
 \end{equation}
 Using the fact that $wZ=w$ \cite{Grinstead2006} and with some calculation one obtains
 
 \begin{equation}
\  L_{j}=\frac{1}{1-w_{j}}\frac{z_{jj}}{w_{j}}\frac{\sum_{i}\sum_{j}d_{ij}^{\beta-\alpha}}{\sum_{i}\sum_{j}d_{ij}^{-\alpha}}+\frac{1}{1-w_{j}}\sum_{k}z_{jk}\left(\frac{\sum_{m}d_{km}^{\beta-\alpha}}{\sum_{m}d_{km}^{-\alpha}}\right).
 \label{19}
 \end{equation}
 Empirically we find that the simulation values of the last term is far less than that of the first term and can be neglected in the analysis. In this context, Eq~.(\ref{19}) reduces to
 \begin{equation}
\  L_{j}\approx{\frac{z_{jj}}{K_{j}}\sum_{i}\sum_{j}d_{ij}^{\beta-\alpha}},
 \label{20}
 \end{equation}
 where $K_{j}=\sum_{m}d_{jm}^{-\alpha}$ named the long-range degree of node $j$ \cite{APRiascos2012}. Here, we omit the value $w_{j}$ as it can be approximated as zero when the network size $N$ is very large. Moreover, for the fractal network with the fractal dimension $d_{f}$, the network diameter $M$ can be approximated as $M\sim{N^{\frac{1}{d_{f}}}}$. Approximating $M$ as a continuous variable, the term $\sum_{i}\sum_{j}d_{ij}^{\beta-\alpha}$ scales as \cite{Li2013}
 \begin{equation}
\
\sum_{i}\sum_{j}d_{ij}^{\beta-\alpha}\sim{N\int_{1}^{M}x^{\beta-\alpha}
x^{d_{f}-1}dx}\sim 
\begin{cases}
  N\frac{N^{\frac{d_{f}+\beta-\alpha}{d_{f}}}-1}{\beta+d_{f}-\alpha}, & \alpha\neq{d_{f}+\beta} \\
   \frac{Nln{N}}{d_{f}}, & \alpha=d_{f}+\beta\\
\end{cases}.
\label{30}
\end{equation}
Plugging Eq.~(\ref{30}) into Eq.~(\ref{20}), we have a linear relationship between $lnL_{j}$ and $\beta$ (i.e., $lnL_{j}\sim{C\beta}$ where $C$ is a constant value determined by the fractal dimension $d_{f}$), when the position of the trapping node $j$ and the tuning exponent $\alpha$ are fixed.

\section*{Acknowledgements}
J.Z. is supported by National Science Foundation of China NSFC 61104143. We thank Kai Zhang for useful discussions and helps.

\section*{Author contributions statement}

T. F. Weng, J. Zhang, M. Small and P. Hui  designed the research, performed the research, and wrote the manuscript. K. Moein and R. Zheng analyzed data and performed research. All authors reviewed the manuscript.

\section*{Additional information}
\textbf{Competing financial interests:} The authors declare no competing financial interests.

\end{document}